\documentclass[twocolumn]{article}
\usepackage{setspace}
\usepackage{graphicx}
\usepackage[super]{natbib}
\usepackage{amssymb, latexsym}
\usepackage{amsmath} 
\usepackage{amsthm}
\usepackage{bm}
\usepackage[mathscr]{eucal}
\usepackage{enumerate}
\usepackage{multirow} 
\usepackage[center]{caption}
\usepackage{musicography}
\usepackage{float}

\usepackage{geometry}

\usepackage{authblk}
\usepackage{abstract}

\begin{document}
	
\title{Neural coincidence detection strategies during perception of multi-pitch musical tones}

\author{Rolf Bader}
\affil{ Institute of Musicology\\ University of
	Hamburg\\ Neue Rabenstr. 13, 20354 Hamburg, Germany\\
}
\date{\today}
	
\twocolumn
[
\begin{@twocolumnfalse}

	\maketitle
	\begin{abstract}
Multi-pitch perception is investigated in a listening test using 30 recordings of musical sounds with two tones played simultaneously, except for two gong sounds with inharmonic overtone spectrum, judging roughness and separateness as the ability to tell the two tones in each recording apart. 13 sounds were from a Western guitar playing all 13 intervals in one octave, the other sounds were mainly from non-Western instruments comparing familiar with unfamiliar instrument sounds for Western listeners. Additionally the sounds were processed in a cochlear model transferring the mechanical basilar membrane motion into neural spikes followed by a post-processing simulating different degrees of coincidence detection. Separateness perception showed a clear distinction between familiar and unfamiliar sounds, while roughness perception did not. By correlating perception with simulation different perception strategies were found. Familiar sounds correlated strongly positive with high degrees of coincidence detection, where only 3-5 periodicities were left, while unfamiliar sounds correlated with low coincidence levels. This corresponds to an attention to pitch and timbre respectively. Additionally, separateness perception shows an opposite correlation between perception and neural correlates between familiar and unfamiliar sounds. This correlates with the perceptional finding of the distinction between familiar and unfamiliar sounds with separateness. 
		\end{abstract}
\end{@twocolumnfalse}
]

\vspace{1cm}

\section{Lead Paragraph}

Brain dynamics during perception is often found to happen in self-organizing neural fields rather than concentrating on single brain areas or even single neurons. Additionally the old notion of subcortical neural structures to be unconsicous is challenged especially in the auditory pathway, where pitch and timbre appear in complex neural fields, and higher cortical brain regions seem to be responsible for associative musical tasks like rhythm extraction, musical tension or music theory. Although the field notion of perception in the auditory pathway works very well with single musical tones, complex musical intervals, chords or polyphonic music will lead to more complex behaviour due to the strong nonlinearity of the cochlear and neural networks. As the auditory pathway is reducing the auditory input through coincidence detection of neurons, we investigate the process of coincidence detection of multi-pitch sounds both in a listening test as well as by cochlear and neural simulations.

\renewcommand \thesection{\Roman{section}}
\renewcommand \thesubsection{\Alph{subsection}} 
\renewcommand \thesubsubsection{\arabic{subsubsection}} 

\section{Introduction}

Multi-pitch sound consists of more than one musical tone, where each tone has a harmonic overtone spectrum constituting musical intervals, chords or polyphonic textures. Helmholtz suggested that in such cases the roughness of the sound increases, and reasoned the musical major scale to consist of such intervals with least roughness\cite{Helmholtz1863}, which are musical fifth, fourth, third, etc. He calculated roughness as musical beating, as amplitude modulation when two partials are close to each other with a maximum roughness at a partial frequency distance of 33 Hz, decreasing to zero when partials align, or when they are spaced wider apart. Other roughness estimation algorithms use similar approaches . Roughness calculations built to estimate industrial noise perception often fail to estimate musical roughness (for a review see \cite{Schneider2018a}).

Another feature of musical sound perception is tonal fusion\cite{Schneider2018b}. A harmonic overtone spectrum consists of many partials or frequencies, still subjects perceive only one musical pitch which is a fusion of all partials. Contrary, inharmonic overtones are clearly separated by perception where the single frequencies can be perceived separately. In a multi-pitch case of musical intervals or chords, trained listeners can clearly distinguish the musical notes, they separate them. Still some multi-pitch sounds are perceived as so dense that separation is not trivial. Therefore a certain degree of perceptual separateness exists, which is a subject of this investigation.

In auditory perception pitch and timbre are clearly distinguished, where Multidimensional Scaling Method (MDS) listening tests showed pitch to be more salient than timbre (for a review see \cite{Bader2013}. This salience of pitch perception was interpreted as the massive presence of the pitch periodicity as interspike-intervals (ISI) in the auditory pathway over all tonotopic auditory neural channels in the presence of a sound with harmonic overtones or partials (in the literature overtones are called the freqeuencies above the fundamental frequency, alternatively a sound consists of partials, where the fundamental is the first partial)\cite{Bader2017}. This is caused by drop-outs of spikes at higher frequencies, therefore constituting subharmonics. All energy-transferring auditory channels constituting a field of neighbouring neurons represent pitch much more prominent than timbre, as the ISI of the fundamental periodicity is present in all such channels, while timbre periodicities only appear in some of them. This might explain the salience of pitch over timbre, which allows melodies or a musical score, mainly representing pitches possible.

Such a field notion of perception and cognition was also suggested for Gestalt perception and the binding problem. Haken uses the mathematical framework of a Synergetic computer to model brain dynamics\cite{Haken2008}. Taking categories as feature vectors, associated with fields of neighbouring neurons, he successfully models completion of only partially present Gestalts. Baars suggests a global workspace dynamics to model levels of perception and binding between all brain regions\cite{Baars2013}. Self-organization was found to be the cause of visual binding in the cat visual cortex\cite{Fries2001}.

Global synchronization of neurons was also found to be the cause for expectancy\cite{Buhusi2005}. Using an Electronic Dance Music (EDM) piece, where by several continuous increases of instrumentation density climaxes are composed which are expected by listeners in this genre, a synchronization of many brain areas was found to peak at the climaxes, decline afterwards, and built up with the next instrumentation density increase\cite{Hartmann2014}.

Perception represented as spatiotemporal patterns appearing in a spatial field of neighbouring neurons have been found in the olfactory, auditory, or visual cortex\cite{Barrie1996}\cite{Kozma2016}. An amplitude modulated (AM) pattern was found, where during the large amplitude phase of AM a spatiotemporal pattern is established in a neural field, which breaks down at the amplitude minimum of the AM. The patterns themselves have been shown to rotate through a large-amplitude cycle.

A similar finding was made in the A1 of the auditory cortex when training gebrils in terms of raising and falling pitches in a discrimination task\cite{Ohl2016}. Using different auditory stimuli and several training phases, a field of neighbouring neurons shows complex patterns. These patterns differ for immediate perception after a few hundred milliseconds and after about 4 seconds, which were interpreted as discrimination and categorization respectively.

Synchronization of auditory input is happening at several stages in the auditory pathway already. Phase alignment of frequencies in a sound with a harmonic overtone spectrum happens already at the transition of mechanical waves on the basilar membrane (BM) to spikes\cite{Bader2015}. Further synchronization is performed in the nucleus cochlearis and the trapezoid body\cite{Joris1994a}\cite{Joris1994b}. Coincidence detection neural models are able to account for spike synchronization to spike bursts up to about 300 - 700 Hz, depending on the neural input strength\cite{Bader2018b}.

Indeed the cochlear seems very effectively tuned to detect mechanisms and useful information on acoustic processes of the outside world. Modern theories of effective auditory coding have been proposed using a cochlea model and correlating its output to a template modeling psychoacoustic data \cite{Dau1996a} \cite{Dau1996b}. Efficient coding was found with the cat auditory nerve compromising between ambient and crackling noise using gammatones \cite{Lewicki2002}. Applying this method to a single sound shows that very few gammatones are needed to code entire sounds \cite{Bader2013}. Mechanisms for the fusion of sound into single pitch have been proposed as autocorrelations of sounds or of common pitch denominators like residual pitch \cite{Cariani2001} \cite{Goldstein1978}. The cochlear is also able to follow acoustical bifurcations on musical instrument sounds, like the surface tone of a cello very precisely\cite{Bader2018a}.

The present study is investigating the perception and neural correlates ot two-tone sounds, consisting of two musical notes plated by the same instrument at the same time. Such sounds will be perceived with a certain amount of roughness and separateness as the ability to hear both tones separately at the same time. These perceived psychoacoustical parameters need to correlate with neural spike patterns in the auditory pathway. Still the perception is expected to follow different strategies depending on the familiarity of musical instrument sounds. With familiar instruments for Western listeners, like e.g. with the guitar, pitch extraction might be an easier task compared to pitch extraction of unfamiliar sounds.

Due to the nonlinearity of the cochlear and the auditory pathway it is not expected that the pitch theory of the fundamental periodicity appearing in all energy-holding nerve fibers due to drop-outs and subharmonics can work the same with two tones as with one tone by simply adding the behaviour of each single tone together. Rather new behaviour is expected and investigated in this paper.

We first describe the method of listening test, spike model and spike post-processing, simulating coincidence detection. Then the results of both methods are discussed. Finally the correlation between perception and calculation is presented. 

\section{Method}

\subsection{Listening Test}

In a listening test 30 musical sounds were presented to 28 subjects aged 19-42 with a mean of 24.9, 15 female and 13 male. They were students at the Institute of Systematic Musicology at the University of Hamburg, all except of 5 played Western musical instruments like guitar, piano,trumpet or drums. There was only one mention of a non-Western instruments, the ukulele. Only five subjects indicated that they were trained in musical interval identification. All sounds presented two musical tones played on real instruments (no samples used), except for two gong sounds, which were played as single gong tones. The gong sounds consist of inharmonic overtone spectra and were used to estimate a multi-pitch perception for percussion instruments. Tab. \ref{table:sounds} gives the instruments used together with the played intervals.

To have a systematic reference of all intervals within an octave, 13 sounds were played with a Western guitar, all with lower pitch g3 and a second pitch above g3 in thirteen half-tone steps from unisono up to the octave g4. The pitches are in the middle range of guitar playing and well within singing range. These pitches were assumed to be familiar to Western listeners.

On the other side, nearly all other sounds were from non-Western instruments like the Chinese \emph{hulusi}, a free-reed wind instrument with three tubes, where two tubes were played. There is no such instrument in the West. Although the harmonica or the accordion are also free-reed instruments, the pitch of these instruments is determined by the reed. With the \emph{hulusi} on the other side the pitch is determined by length of a bamboo tube the reed is attached to. Using regular fingering technique different pitches can be played. The \emph{hulusi} used was collected by the author in Lijiang, Yunnan, China.

The \emph{saung gauk} is a Burmese arched harp. Its strings are attached to an arched stick on both sides, where one side of the stick is attached to a resonance box covered with ox leather, which again is covered with heavy lacquer. It is astonishing loud for a harp, which is caused by resonances of the stick and the resonance body. For the experiment a Mong crocodile \emph{saung gauk} was used, collected by the author in Yangon, Myanmar.

The \emph{roneat deik} used in the experiment is a metallophone consisting of 21 metal plates collected by the author in Phnom Penh, Cambodia. Although gong ensembles are common in Southeast Asia, the \emph{roneat deik} is the seldom case of a metal instrument, most other are made of bronze. The sound of each plate has an inharmonic overtone spectrum, still its fundamental is strong, making it possible to play melodies with it. Still it is expected that the sound of the instrument is unfamiliar to Western listeners.

The \emph{dutar} used is a long-neck lute collected by the author in Kashgar, Xinjiang, China. It is played as a folk instrument by the Uyghur people along the traditional silk road. It has two strings, where two pitches are played at the same time. Although it is a plucked string instruments like the guitar, the sound is considerably different from a guitar or a lute, caused by the unusual length of the strings on a long-necked lute and the small resonance body.

The \emph{mbira} is also known as thumb piano. It consists of 9 metal plates attached to a wooden resonance box. Although the spectrum of the metal rods are inharmonic, the fundamental frequencies of each spectrum is much stronger in amplitude than the higher partials, and therefore melodies can be played on the instrument. Its origin is unkn
own, but it was most likely built for Western musicians in the World Music scene.

The two gongs are collected by the author in Yangon, Myanmar, and are played within the \emph{hsain waing} ensemble of traditional Bama music. The large gong has a diameter of 54cm, the diameter of the small one is 26cm.


\begin{center}
\begin{table}[H]
\begin{tabular}{|c|c|c|}
	\hline 
	Nr. & Instrument & Pitches \\ 
	\hline 
	1 & Western guitar & Unisono, g3- g3 \\ 
	\hline 
	2 & Western guitar & Minor Second, g3-g\#3 \\ 
	\hline 
	3 & Western guitar & Major Second, g3-a3\\ 
	\hline 
	4 & Western guitar & Minor Third, g3-b flat{}3\\ 
	\hline 
	5 & Western guitar & Major Third, g3-b3 \\ 
	\hline 
	6 & Western guitar & Fourth, g3-c4  \\ 
	\hline 
	7 & Western guitar & Tritone, g3-c\#4 \\ 
	\hline 
	8 & Western guitar & Fifth, g3-d4  \\ 
	\hline 
	9 & Western guitar & Minor Sixth, g3-e flat{}4  \\ 
	\hline 
	10 & Western guitar & Major Sixth, g3-e4 \\ 
	\hline 
	11 & Western guitar & Minor Seventh, g3-f4  \\ 
	\hline 
	12 & Western guitar & Major Seventh, g3-f\#4 \\ 
	\hline 
	13 & Western guitar & Octave, g3-g4  \\ 
	\hline 
	14 & Saung Gauk & Minor Seventh, f\#3-e4 \\ 
	\hline 
	15 & Saung Gauk & Minor Sixth, f\#3-d4 \\ 
	\hline 
	16 & Saung Gauk & Major Second, f\#3-g\#3  \\ 
	\hline 
	17 & Dutar & Major Seventh, f\#2-e3 \\ 
	\hline 
	18 & Hulusi & Major Third, c4-e4 \\ 
	\hline 
	19 & Hulusi & Fifth, a3-e4  \\ 
	\hline 
	20 & Hulusi & Minor Third, e4-g4  \\ 
	\hline 
	21 & Hulusi & Minor Seventh, f\#3-e4 \\ 
	\hline 
	22 & Mbira & Minor Third, c4-e4 \\ 
	\hline 
	23 & Bama Big Gong & Fundamental $\sim$ b2 \\ 
	\hline 
	24 & Bama Small Gong & Fundamental $\sim$ g\#2 \\ 
	\hline 
	25 & Roneat Deik & Major Second, f\#4-g4  \\ 
	\hline 
	26 & Ronest Deik & Two octaves plus\\
	& &  major second, b3-c6\\ 
	\hline 
	27 & String Pad & MinorSecond, d\#4-e4 \\ 
	\hline 
	28 & String Pad & Four octaves plus \\ 
	& &Major Third, f\#2-a\#6\\
	\hline 
	29 & Piano & Fifth, c4-g4 \\ 
	\hline 
	30 & String Pade & Minor Seventh, c2-b flat{}2 \\
	\hline 
\end{tabular} 
	\caption{List of sounds used in analysis and listening test. The guitar sounds cover the whole range of half-tone steps in one octave. They are supposed to be familiar for Western listeners. The other sounds are nearly all from non-Western instruments, and therefore taken to be unfamiliar for Western listeners.}

\label{table:sounds}
	\end{table}
	\end{center}

Two other sounds have been used, a piano and a synth pad sound. These sounds were created by a Kawai MP-13 keyboard. The pad sound was used as a continuous sound, which most of the other sounds are not. The piano sound was used because it is a standard instrument for musical training. Pad sounds are generally familiar to Western listeners. Still as synth sounds can be composed by any synthesis method with any parameters, these sounds were chosen to be unfamiliar. Also the piano sounds were chosen to be added to the unfamiliar sounds in the analysis. Although this is not perfectly correct, the advantage is to keep the 13 guitar sounds as a group of homogeneous sounds. There are only two piano sounds in the sound set and taking them out does not change the results considerably.

Subjects were presented all sound three times and asked to rate their perception of the sounds roughness and separateness on a scale of 1-9 each, where 1 is no roughness / no separateness and 9 is maximum roughness / maximum separateness. To give subjects an idea about where these maxima might be before the test several sounds used in the test were played which in a pretest turned out to be those perceived most rough and separate respectively.

Additionally the subjects were asked to identify the musical interval of the sounds, e.g. fifth, fourth, etc. This task was optional, as it was not clear how many subjects would have the capability of identifying the intervals and those with less capacity were asked not to wast time and attention to this task. A small questionnaire asking for age, gender and musical instrument played by subjects was presented before the listening test. There also musical interval training of subjects was asked for, where only five subjects indicated they had intense training. Indeed only five subjects answered some of the intervals, most of them were wrong. Therefore in the follow it is not possible to correlate correct interval identification by subjects with the other findings. Still we can state that nearly all subjects did not have a considerable musical interval identification capability.

\subsection{Cochlear Model}

All sounds were processed using a cochlear model and post-processing the cochlear output. The cochlear model was used before in a study of phase-synchronization of partials in the transition between mechanical motion on the basilar membrane into neural spikes \cite{Bader2015}, in a coincidence detection model using a Iszikevich neural model \cite{Bader2018b}, when studying cello surface sounds \cite{Bader2018a}, or as foundation of a pitch theory \cite{Bader2017}.

The model assumes the basilar membrane (BM) to consist of 48 nodal points and uses a Finite-Difference Time-Domain (FDTD) solution with a sample frequency of 96 kHz to model the BM motion. The BM is driven by the ear channel pressure, which is assumed to act instantaneously on the whole BM. This is justified as the sound in the peri- and endolymph is about 1500 m/s, where the sound speed on the BM is only 100 m/s down to 10 m/s. The pressure acting on the BM is the input sound. The output of the BM is a spike activity, where a spike is released at a maximum of the positive slope of the BM and a maximum slope in the temporal movement at one point on the BM. This results in a spike train similar to measurements. 

With periodic sounds the spikes are only present at the respective Bark bands of best frequency of the periodic sound. With harmonic overtone spectra all Bark bands of the spectral components show spikes, while the other Bark bands do not. Also the delay between the high frequency spikes to the low frequencies at the BM apex is about 3-4 ms, consistent with experiments. The model does not add additional noise. It also has a discretization of 48 nodal points, which is two nodal points for each Bark band.

\subsection{Post-processing of Cochlear Model Output}

The output of the cochlear model are spikes at certain time points along the BM in the Bark bands. From here the interspike intervals (ISI) are calculated as the time interval between two spikes at one nodal point, so two for each Bark band. Accumulating the ISI over a time interval and transforming the ISI into frequencies like f=1/ISI, an ISI histogram of the amount of appearance or the amplitude of certain frequencies or periodicities present in all Bark bands results. The accumulation interval was chosen according to logarithmic pitch perception to be 10 cent, where one octave is 1200 cent.

To simulate the coincidence detection found in neural nuclei subsequent to the cochlear, the ISI histograms as shown above are post-processed in three stages. First the histograms of all time steps are summed up to build one single histogram for each sound. As shown in Fig. \ref{fig:ISIHistogram_ExampleSounds_1} and Fig. \ref{fig:ISIHistogram_ExampleSounds_2} the histograms change over time, mainly due to the temporal decay of the sound. They also show inharmonicities during the initial transient. After temporal integration the ISI histogram is divided by the amplitude of its largest peak, making the amplitude of this peak one. A threshold of th=0.01 is applied, making all peaks smaller than th zero. Additionally, all frequencies above 4kHz are not taken into consideration, as pitch perception is only present up to this upper frequency.

Secondly, a Gauss blurring of the spectrum is performed, where the spectrum is blurred using a Gauss shape with standard deviation $\sigma$. This sums peaks which are very close to each other, and therefore performs coincidence detection of periodicities close to each other. In a second step only peaks with a certain sharpness s are used, where s=0 selects all peaks and s>0 selects only peaks, where the negative second derivative of the spectrum is larger than s. This prohibits that after blurring the spectrum very broad peaks are still taken as such peaks. To estimate the amount of coincidence detection in the listening test both values are varied in 21 steps between $0 \leq \sigma \leq 2$, $0 \leq s \leq 0.002$.

In a third step the ISI histogram is fused into a single number to make it comparable to the judgments of separateness and roughness of the listening tests, which are also single numbers for each sound. Three methods were used, the amount of peaks N, a sum of the peaks, each weighted with its amplitude like

\begin{equation}
W = \sum_{i=1}^N ISI_i \ ,
\end{equation}

and a Shannon entropy like

\begin{equation}
S = 1/\log{N} \sum_{i=1}^N ISI_i \log{ISI_i} \ ,
\end{equation}

where N is the amount of peaks and $ISI_i$ is the amplitude of the i$^{th}$ peak.

The three methods decide which perception process is more likely to occur in listeners when correlating the results with the separateness and roughness perception.

\subsection{Correlation between perception and calculation}

In a last step the mean values for the perception of separateness and roughness respectively over all subjects were correlated with the results of the simulation. To address the hypothesis of different perception strategies for familiar, in this case Western and unfamiliar instruments, the correlations were performed for two subsets of the 30 instruments. One subset consists of all 13 guitar sounds, the other subset are all other sounds. The 13 guitar sounds form a consistent set, including all 13 intervals in the octave of a Western scale. The non-Western instruments have different intervals, the two gongs consist of inharmonic spectra for comparison. In the follow the terms Western and guitar sounds, as well as non-Western and non-guitar sounds as used synonymously.

So for each perception parameter, separateness and roughness, twelve cases exist, three summing methods N, W, and S, each for mean and standard deviation of the perception parameters, and each for the Western and non-Western subsets. Each of these twelve cases consist of 21 $\times$ 21 = 441 correlations for all combinations of $\sigma$ and s used. For the sake of clarity, the combinations of $\sigma$ and s are combined in one plot.

\section{Results}

\subsection{ISI histogram}

\begin{figure}[H]
	\centering
	\includegraphics[width=0.7\linewidth]{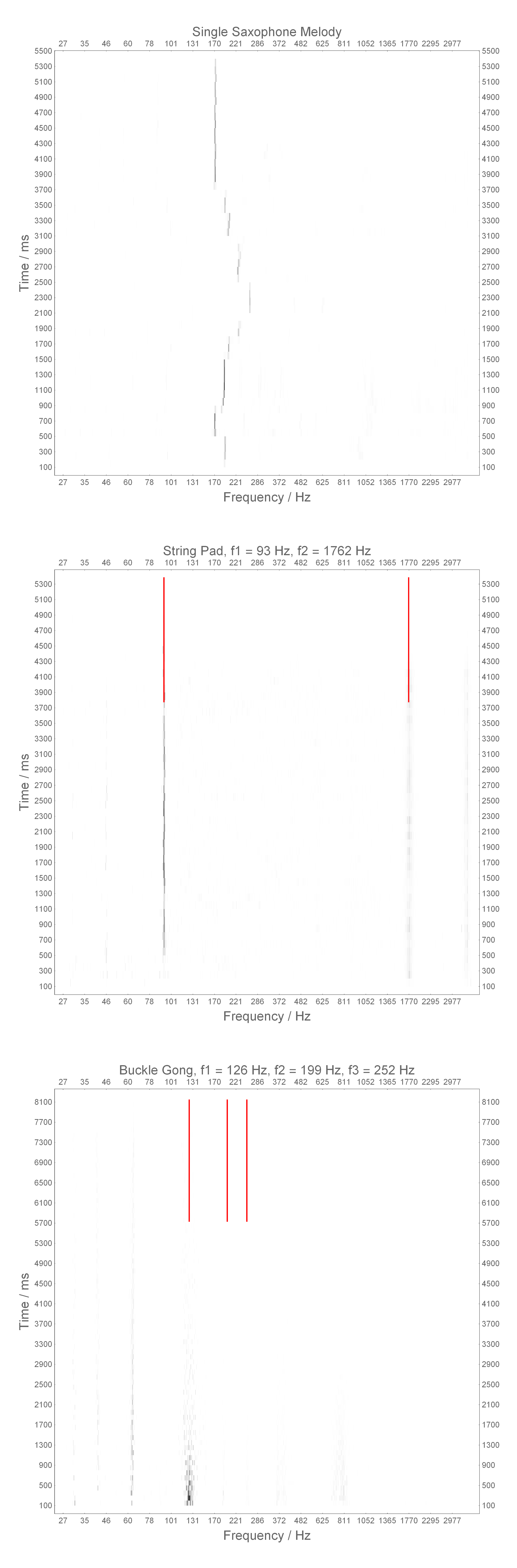}
	\caption{Temporal development of ISI histogram, summing all ISI in all Bark bands and transferring them into frequencies f = 1 / ISI for three example sounds: top: a single tone saxophone melody, middle: a string pad sound consisting of two notes with fundamental frequencies $f_1 = 93$ Hz and $f_2 = 1762$ Hz, bottom: A large buckle gong with three lowest frequencies $f_1 = 126$ Hz, $f_2 = 199$ Hz, and $f_3 = 252$ Hz. The lines at the upper part of the middle and bottom plots indicate these frequencies. The saxophone melody is perfectly represented by the strongest peak at all ISI histograms, as shown in Fig. \ref{fig:Jazz_Melodien_f0_Autokorr_vs_CochlearModel} in detail. The string pad sounds fundamentals are also clearly represented by the ISI histogram. The gongs lowest partial is clearly there, still a residual ISI below this $f_1$ is there, too.}
		\label{fig:ISIHistogram_ExampleSounds_1}
	\end{figure}

\begin{figure}[H]
	\centering
	\includegraphics[width=0.7\linewidth]{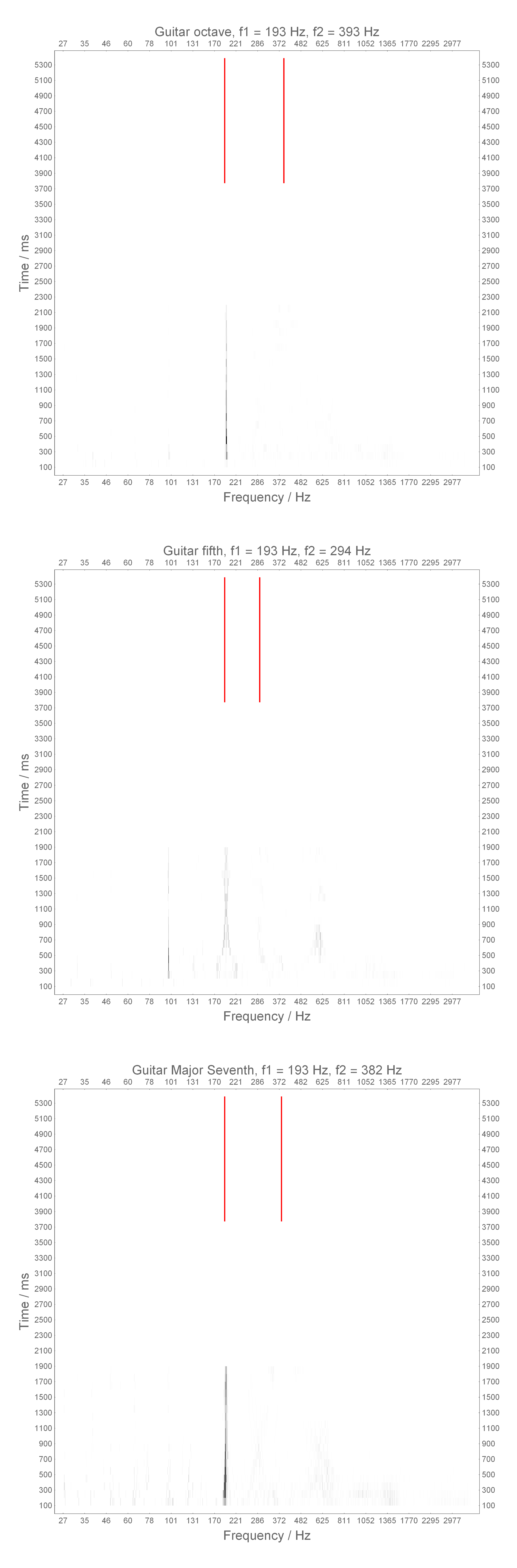}
	\caption{Same as Fig. \ref{fig:ISIHistogram_ExampleSounds_1} for three two-note intervals played on a classical guitar. All have the same fundamental pitch g at $f_1 = 193$ Hz. Top: octave above $f_1$ at $f_2 = 393$ Hz, middle: fifth above $f_1$ at $f_2 = 294$ Hz, bottom: major seventh above $f_1$ at $f_2 = 382$ Hz. In all cases $f_1$ is clearly represented in the ISI histogram. With the octave the $f_2$ partial is nearly not present, with the fifth $f_2$ is again only slightly there, still a residual periodicity below $f_1$ appears. With the major seventh again $f_2$ is nearly not present, still many periodicities above and below $f_1$ are slightly present.}
	\label{fig:ISIHistogram_ExampleSounds_2}
\end{figure}

Fig. \ref{fig:ISIHistogram_ExampleSounds_1} and Fig. \ref{fig:ISIHistogram_ExampleSounds_2} show six examples of ISI histograms, each consisting of adjacent time intervals, where each interval is an integration over the cochlear spike output in this interval. Five of these six examples have been used in the listening test, only the top plot of Fig. \ref{fig:ISIHistogram_ExampleSounds_1} is presented as a single-note example to exemplify the pitch theory and the salience of pitch over timbre discussed above.

\begin{figure}[H]
	\centering
	\includegraphics[width=1\linewidth]{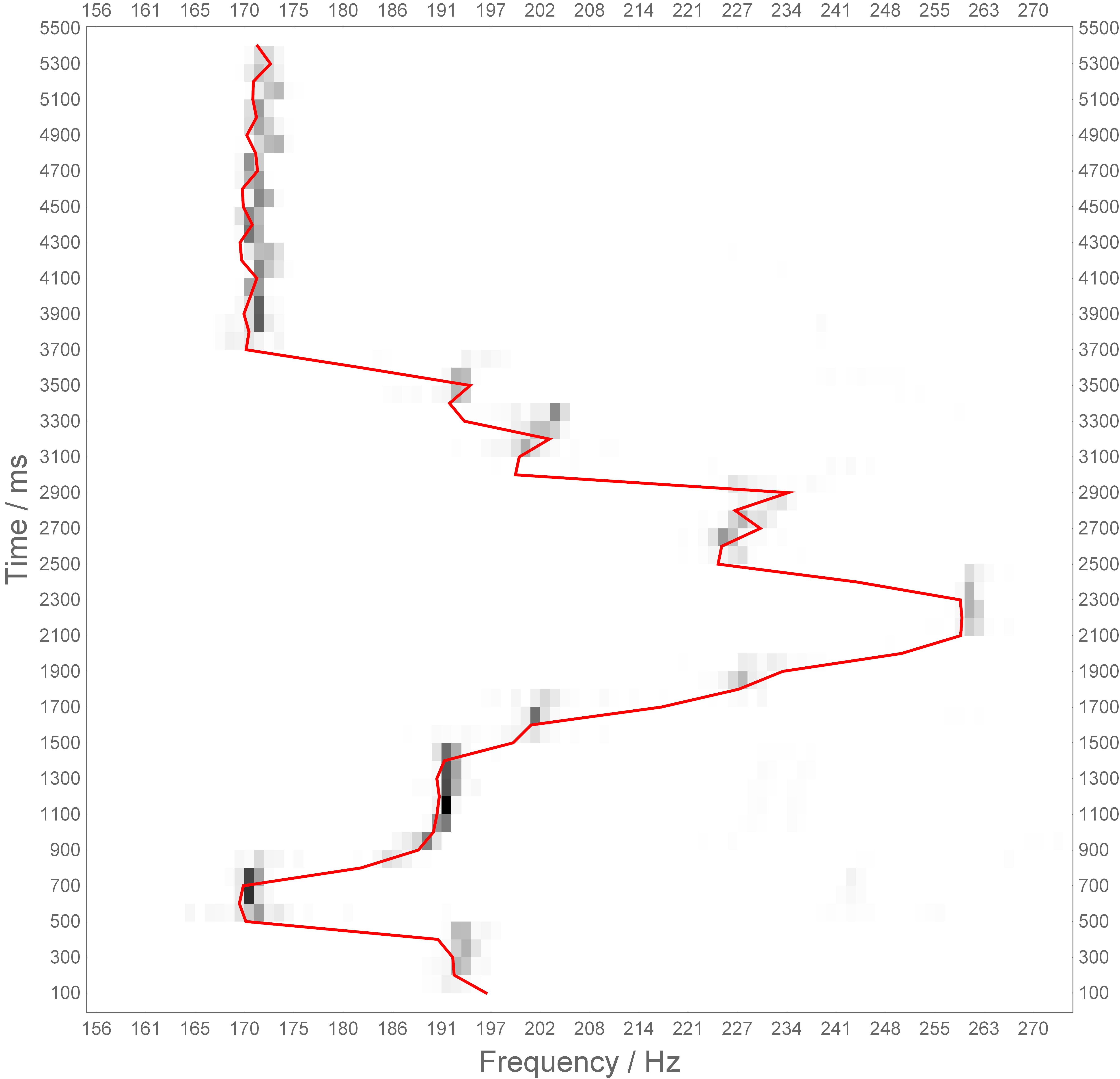}
	\caption{Excerpt from Fig. \ref{fig:ISIHistogram_ExampleSounds_1} top of the single saxophone line, comparing the ISI histogram temporal development with the results of an autocorrelation function of the sounds time series when using the first peak of the autocorrelation as $f_0$ at adjacent time intervals. Both curves align very well. Therefore the strongest peak of the ISI histogram is representing the fundamental pitch very precisely.}
	\label{fig:Jazz_Melodien_f0_Autokorr_vs_CochlearModel}
\end{figure}

This single-note example is a single-line saxophone melody. It appears that at all time intervals there is only one strong peak. Aurally this peak follows the played melody very well. To verify this perception, in Fig. \ref{fig:Jazz_Melodien_f0_Autokorr_vs_CochlearModel} an excerpt of Fig. \ref{fig:ISIHistogram_ExampleSounds_1} top plot is shown. It shows the whole time but is restricted to the area of large peaks. The continuous curve overlaying the ISI histogram plot is the result of an autocorrelation calculation using the plain sound time series as input. With 50 ms intervals the autocorrelation of the sound time series is calculated, and the interval between the autocorrelation beginning and the first autocorrelation function peak is taken as fundamental periodicity. This is an established method for detecting $f_0$ in single melodies, and has e.g. be applied to analyze intonation of Cambodian Buddhist \emph{smot} chanting \cite{Bader2011}. Both functions, the autocorrelation and the ISI histogram align very well. Therefore the largest peak in the ISI histogram can be used for $f_0$ extraction in single melody lines.

Still we are interested in multi-pitched perception. Therefore a second example is that of a synthetic sound, a string pad, playing a two-note interval of four octaves and a major third (sound 28, f\#2-a\#6) with two fundamental frequencies at $f_1 = 93$ Hz and $f_2 = 1762$ Hz. This sound is particularly interesting, as both pitches were perceived most separate (see results of listening test below). Indeed the two pitches are clearly represented as major peaks in the ISI histogram. In this plot and the following the expected peaks are indicated by solid lines starting from the top down to .7 of the plots hight. The precise frequencies were calculated using a Morelet Wavelet transform of the original sounds.

The third example is a Myanmar large buckle gong (sound 23). Its fundamental frequency is at $f_1 = 126$ Hz. Additionally the next two partials at $f_2 = 199$ Hz and $f_3 = 252$ Hz are indicated in the plot in Fig. \ref{fig:ISIHistogram_ExampleSounds_1} middle plot too. Such a gong has a basically inharmonic spectrum, still many percussion instruments are carved such that the overtone spectrum is tuned with the gong. In an interview with a gong builder in Yangon during a field trip, the instrument maker explained that he is hammering the gong in such a way to tune the third partial to the fifth of the fundamental.

Indeed the gong is tuned very precisely. The relation $f_3 / f_2 = 2$ is a perfect octave, $f_2 / f_1 = 1.58$ is nearly a perfect fifth. In the ISI histogram the lowest partial $f_1$ is the strongest periodicity, $f_2$ and $f_3$ are nearly not present. Still there is a residual around 63 Hz, the octave below $f_1$, and two lighter ones two and three octaves below. Indeed we would expect a residual pitch to appear around 63 Hz, which is actually there. The Wavelet transform does not show any energy below $f_1$ of this gong. Therefore this gong is similar to Western church bells, where the hum note is a residual\cite{Lau2009}. The ISI histogram at $f_1$ is also blurred, which most probably is caused by the beating of the gong, which is not perfectly symmetric causing degenerated modes.

Three other examples are shown in Fig. \ref{fig:ISIHistogram_ExampleSounds_2}. All are two-note intervals played on a classical guitar and all have the same low note g3 with a lowest partial at 193 Hz (called $f_1$ here to compare with the second note with lowest partial at $f_2$). Three intervals are shown, the octave at the top, a fifth in the middle, and a major seventh at the bottom. Again the fundamental frequencies of the two notes were also determined using a Wavelet transform of the original sound time series, and indicated with horizontal lines starting at the top of the plots.

With the octave sound only one peak at $f_1$, the fundamental frequency of the low note appears. The fifth is more complicated, there is energy at $f_1$, which is bifurcating at first only to join after about 800 ms. There is nearly no peak at the fundamental frequency of the second note, the fifth. Still the residual pitch expected at 101 Hz ($f_2 - f_1$) is clearly present. With the major seventh interval as the bottom plot in Fig. \ref{fig:ISIHistogram_ExampleSounds_2} $f_1$, the fundamental of the low note is very strong. But there is nearly no energy at the fundamental of $f_2$, the major seventh. Still many small peaks are present, especially during the first about 500 ms.
	
It has been shown previously, that with periodic sounds, in the Bark band of the fundamental of the sound, ISI are present in all Bark bands with energy, not only at the Bark band of the fundamental frequency. The reason for this is straightforward. A periodic wave form has regions of strong and weak amplitudes during one period. During regions of weak amplitude not much energy is entering the cochlear, and therefore there is not much energy to trigger a nervous spike. So we expect drop-outs in the series of spike in higher frequencies. These drop-outs are experimentally found and also appear in the present cochlear model. These drop-outs are subharmonics of the frequency of the Bark bands, where the common subharmonic to all Bark bands with energy is the fundamental frequency of the harmonic input sound.
	
This means that the fundamental periodicity is present in all Bark bands and is so transferred to higher neural nuclei. This leads to the suggestions of a pitch perception, which is caused by the magnitude of the pitch periodicities present in the whole field of the nervous system carrying these spikes. The strength and salience of pitch over timbre, as found in many Multidimensional Scaling Method (MDS) experiments and which makes melody and score representation of sound with pitches possible at all, might be caused by the salience of the fundamental periodicity in the auditory pathway.
	
\subsection{Perception of Separateness and Roughness}
	
\begin{figure}[H]
	\centering
	\includegraphics[width=1\linewidth]{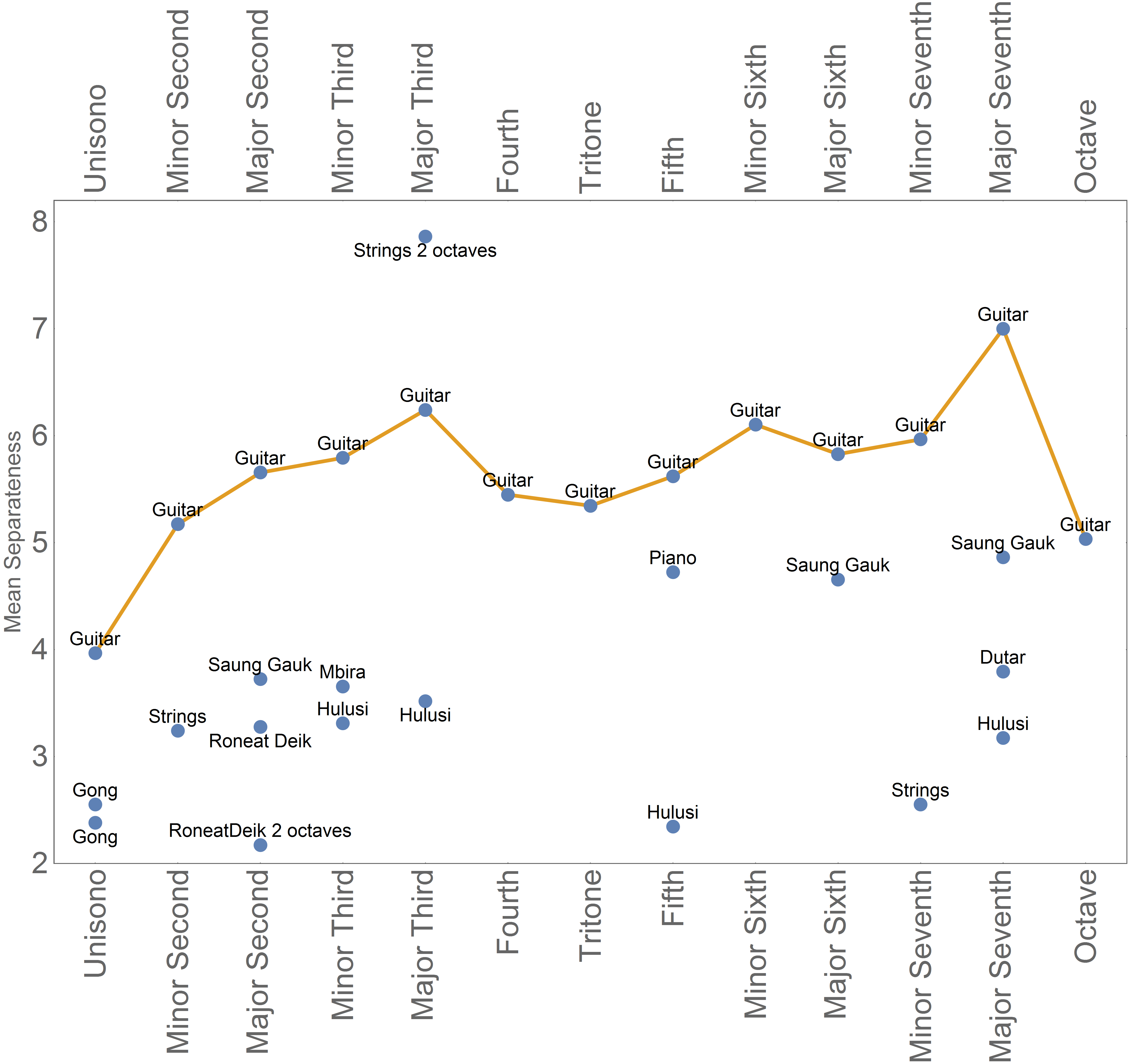}
	\caption{Mean perception of separateness of all stimuli sorted according their musical interval (gongs are sorted as unisono). As the guitar tones form the familiar subgroup and is the only instrument having all 13 interval in the octave, they are connected with the yellow line. Except for the 'Strings 2 octaves' sound, which is the string pad sound, which ISI histogram is shown in Fig. \ref{fig:ISIHistogram_ExampleSounds_1}, all non-familiar sounds are perceived as well separate and therefore as more fused compared to the familiar guitar sound. This points to a different perception strategy of familiar and unfamiliar sounds. Furthermore, the guitar sounds show least separateness in unisono and octave as expected, and most separateness with the major seventh.}
	\label{fig:perceptionseparatenessmean}
\end{figure}

\begin{figure}[H]
	\centering
	\includegraphics[width=1\linewidth]{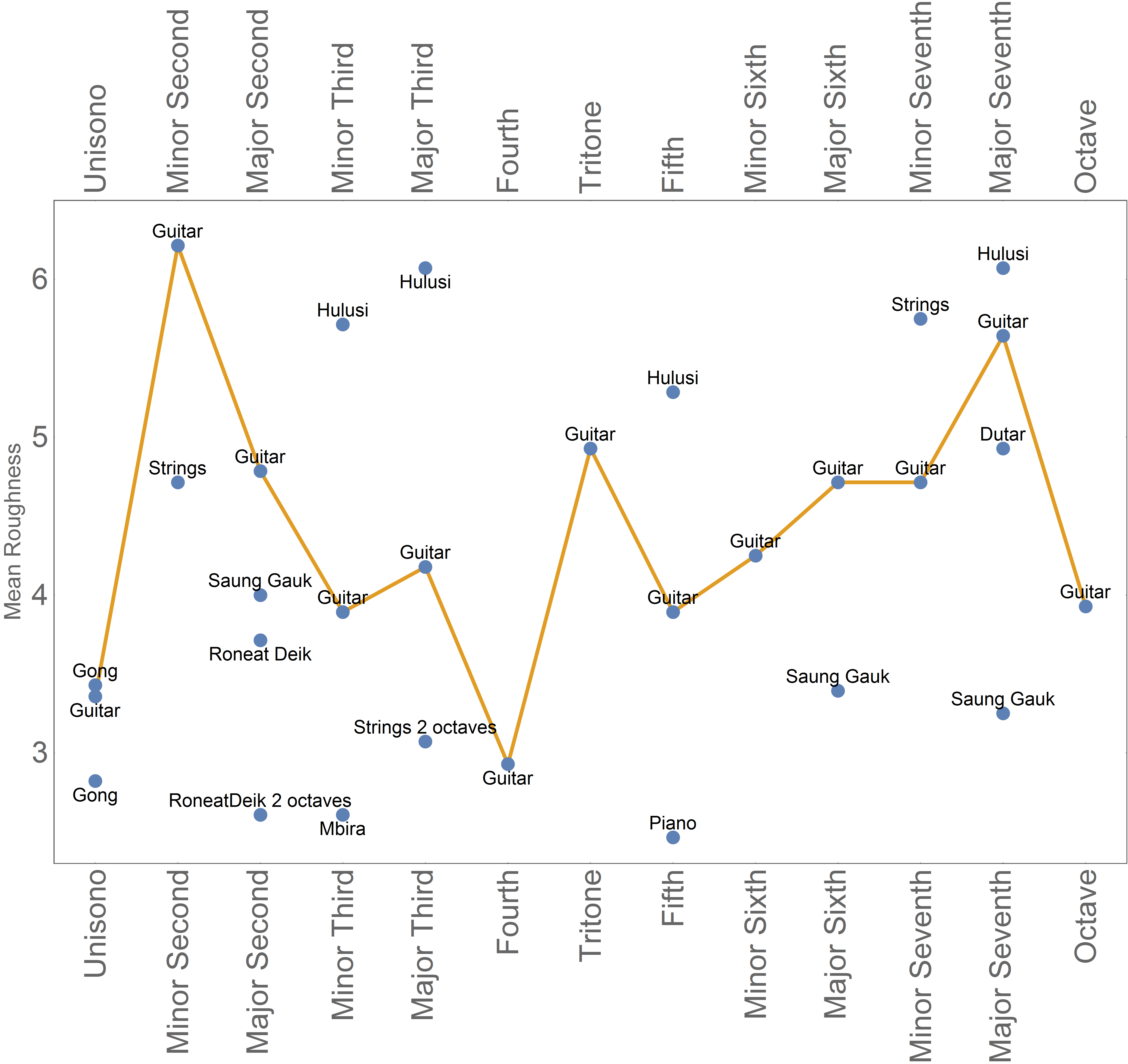}
	\caption{Mean roughness perception of all stimuli, again the guitar tones are connected with the yellow line. The roughness perception of the guitar sounds are as expected, least with unisono, octave, fourth and fifth, and strongest with minor second, major seventh, and tritone. Still the non-familiar instruments are distributed above and below the guitar sounds, and do not show such a simple split than with separateness perception.}
	\label{fig:perceptionroughnessmean}
\end{figure}

The mean perception over all subjects of separateness is shown in Fig. \ref{fig:perceptionseparatenessmean}, the mean perception of roughness in Fig. \ref{fig:perceptionroughnessmean} sorted by the musical interval present in the sound. The familiar guitar sound subset include all intervals, and to make this subset clearly visible the guitar tones are connected with the yellow line.
	
The separateness perception shows nearly all unfamiliar sounds perceived less separated than the guitar sounds. This points to different perception strategies between these two subsets, familiar and unfamiliar sounds. The only exception is the 'Strings 2 octaves' sound which is shown with its ISI histogram in Fig. \ref{fig:ISIHistogram_ExampleSounds_1} middle plot. Here the two notes are separated by over an octave and the ISI histogram shows clearly two distinguished peaks at the respective fundamental frequencies of the two notes in this sound.mple split than with separateness perception. 

	When discussing instrument families the \emph{dutar} and the \emph{saung gauk} are both plucked stringed instruments like the guitar. Still both are lower than the guitar tones in terms of separateness perception. Still this also holds for the two strings sounds, which are also supposed to be familiar with Western listeners and which are also perceived more fused than separated. Still these sounds are artificial string pad sounds, which have above already been discussed as unfamiliar insofar, as such synth sounds can be produced with arbitrary synthesis methods and parameters, which make these sounds unpredictable and therefore unfamiliar to listeners too. The only sound which is still familiar would then be the piano, which is indeed close to the guitar sound in terms of separateness. The least separated sounds are the gongs and the \emph{roneat deik}, which are percussion instruments in the sense of an inharmonic overtone spectrum, although the \emph{roneak deik} is a pitched instrument. With the four instruments having a major seventh interval it is interesting to see how strong the influence of the instrument sound is in terms of separateness perception.

The roughness perception shown in Fig. \ref{fig:perceptionroughnessmean} does not show such a clear distinction between familiar and unfamiliar sounds. The guitar tones show a roughness perception as expected, least with the unisono, octave, fourth, and fifth, and strongest with the minor second, major seventh, and tritone. It is interesting to see that all four \emph{hulusi} sounds are above the guitar, so perceived more rough, following the roughness perception of the guitar qualitatively (major seventh most rough, fifth least rough, etc.). 

Also the gongs and the \emph{roneat deik} are perceived with low roughness, they have also been perceived as low in separateness. Still overall no such perception strategy can be found with roughness as it clearly appears with separateness perception.

	Comparing the guitar perception of roughness and separateness is appears that they are counteracting up to the fifth and aligning from the fifth to the octave qualitatively. So above the fifth it is likely for familiar sounds to be perceived separate when they are also perceived rough. Below the fifth it is likely that these two perceptions are opposite. Therefore the perception parameter of separateness is a different perception than that of roughness.
	
		\subsection{Correlation between perception and simulation}
	
	In the section above we have found a clear split between familiar and unfamiliar sounds, pointing to two different perception strategies. Therefore in this section correlating perception to simulation, the two subgroups, familiar guitar sound and unfamiliar sounds are correlated separately with simulation.

\begin{figure}[H]
		\centering
		\includegraphics[width=1\linewidth]{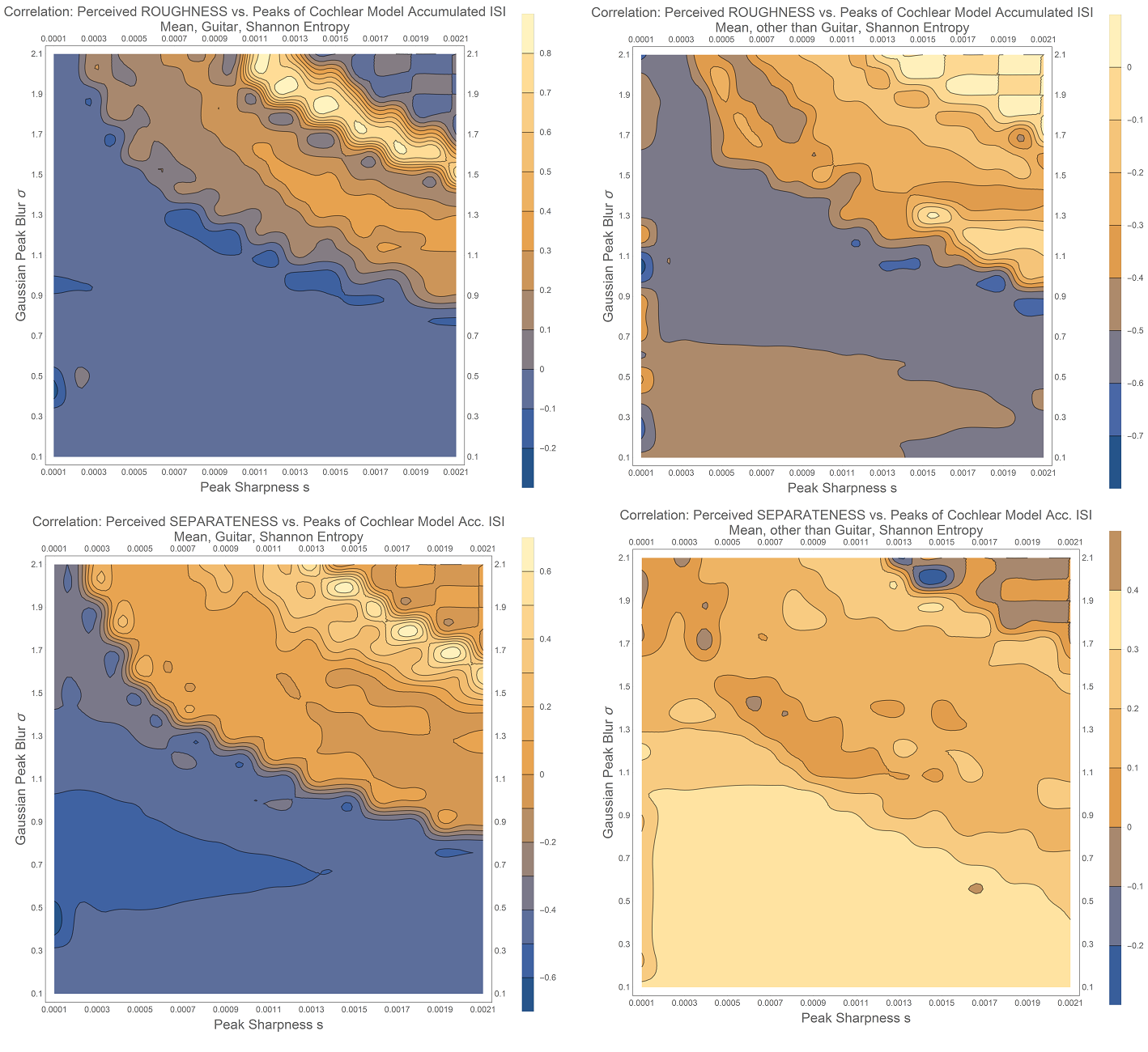}
		\caption{Correlation between perceived roughness (upper plots) as well as separateness (lower plots) for the guitar tones (left plots) and all other no-guitar sounds (right plots) and the entropy S of the accumulated ISI spectrum, while post-processing the spectrum with a Gaussian blurring with standard deviation $\sigma$ and a peak sharpness s. All plots show a consistency or ridge from right/bottom to left/top as expected (see text for more detail). For roughness the no-guitar sounds on the right have a negative correlation throughout, while the guitar sounds on the left have a positive correlation, pointing to an opposite perception between familiar and unfamiliar sounds. For separateness the correlation between the guitar sounds on the left is positive at the right upper ridge and negative in the lower left side, while the non-guitar sound on the left have a positive correlation on this lower left side and negative correlation at the upper right ridge, again showing opposite behaviour.}
		\label{fig:correntropymean}
	\end{figure}

	Fig. \ref{fig:correntropymean} shows the correlation between simulation and separateness as well as roughness perception for the guitar and non-guitar sounds using entropy S. The upper graphs are roughness, the lower two separateness correlations, the left plots are the guitar and the right plots the non-guitar sounds respectively. Each plot varies the standard deviation $\sigma$ of the ISI histogram blurring from $0 \leq \sigma \leq 2$ and a peak sharpness s from $0 \leq s \leq .002$ in 21 steps each (for details see above). Therefore the very left lower corner of each plot is the unprocessed ISI histogram which might contain many peaks, most of them with low amplitude. With increasing $\sigma$ a coincidence detection is performed, summing neighbouring peaks to a single peak. This reduces the amount of peaks, but blurs them. When increasing s only sharp peaks are used. Of course blurred peaks can only appear with higher values of $\sigma$. Therefore higher values of s prevent that for large values of $\sigma$ peaks are detected which are so broad that they cannot reasonably have been built in a coincidence detection process as they are too far apart.

Examining the plots in Fig. \ref{fig:correntropymean}, concentrating on the left upper plot of roughness correlated with guitar tones, a ridge of positive correlation appears in the upper right corner. The rest of the plot is structured accordingly with a tendency of lower right to upper left ridges. Following the ridge with the highest correlation starting at $\sigma \sim 1.5$ on maximum s and ending at s = 0.0011 with maximum $\sigma$, the trade-off between $\sigma$ and s leads to a reduction of the amount of amplitudes down to 3-5 amplitudes in the post-processed ISI histogram only. This is a tremendous reduction taking into consideration that the amount of amplitudes in the unprocessed lower left corner of the plots is between 700-800 amplitudes. On the right lower side of the ridge with strong positive correlation the peaks are not blurred to a maximum as sharpness is very rigid, allowing only a few amplitudes. On the left upper side of the ridge the blurring is increased, still the sharpness criterion is eased, resulting in about the same amount of amplitudes, 3-5. Indeed in the upper right corner only one or even no amplitudes have survived.	

	Again examining this plot, the ridge shows a rippling, as does the region in the right upper corner. These are artifacts of the visualization which uses an interpolation to smooth the plot. Without such interpolation the ridge would consist of single points on a grid which is not different in content but more more unpleasant visually. The visualization of the plots have been varied to much extend, and the present version was considered the best choice. So e.g. the range of the plots have been kept individual to allow for maximum differentiation within the plots. Still this makes the comparison of plots more complex, as coloring of two neighbouring plots correspond to different absolute values. Other color options have been tried too, like absolute coloring, still again, as there are plots of low overall range many colors need to be used making plots with a wide range unreadable.

\begin{figure}[H]
		\centering
		\includegraphics[width=1\linewidth]{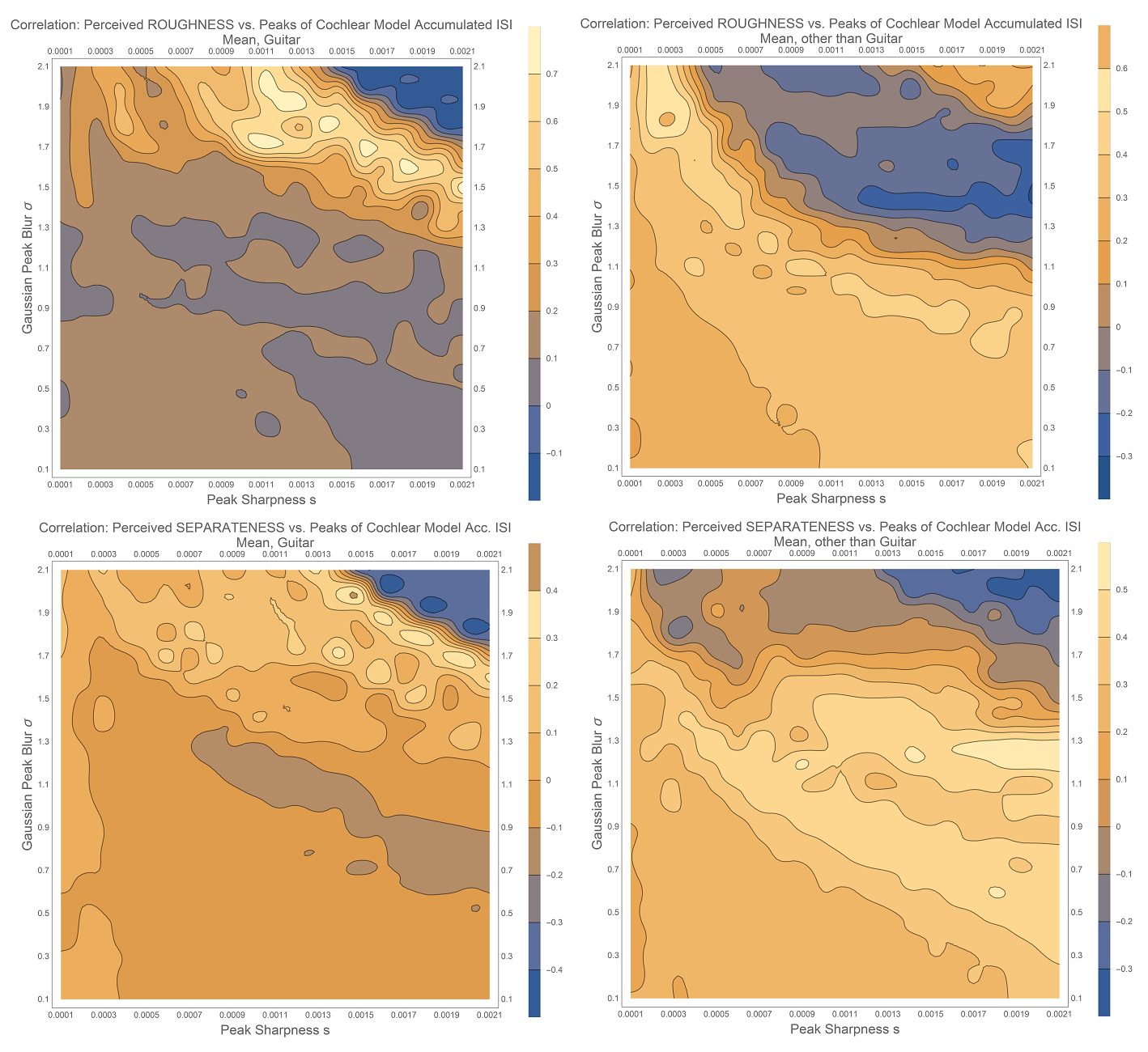}
		\caption{Same as Fig. \ref{fig:correntropymean} but correlation over number of ISI histogram amplitudes N. For roughness the no-guitar sounds on the right have a negative correlation in the upper right part and positive the the lower left, while the guitar sounds on the left have a positive correlation on the upper right ridge, again pointing to an opposite perception between familiar and unfamiliar sounds. For separateness the correlation between the no-guitar sounds on the right is negative at the upper right part and positive the lower left, while for the guitar sounds is positive at the upper right ridge and slightly negative at the lower left part. Again this is an opposite behaviour, although not that clear as with the entropy S in Fig. \ref{fig:correntropymean}.}
		\label{fig:corrmean}
	\end{figure}

	Now comparing the two roughness plots for guitar (left upper) and non-guitar (right upper) sounds in Fig. \ref{fig:correntropymean} they show opposite correlations. While the ridge in the guitar plot has a positive correlation, the correlation at the ridge position with the non-guitar tones is slightly negative, this plot shows negative correlations throughout. A similar behaviour holds for the separateness perception in the lower two plots of this figure. On the left the guitar plot has a positive correlation at the ridge, while the non-guitar tones have a slightly negative correlation there, and is positive in the lower left part. Still there is a difference between separateness and roughness. While the roughness plot is negative in the lower right corner of the plots for both, guitar and non-guitar tones, the separateness plots show negative correlations with the guitar and positive ones with the non-guitar tones. In both cases, roughness and separateness, the ridge is positive with the guitar tones and slightly negative with the non-guitar sounds.
	
	\begin{figure}[H]
		\centering
		\includegraphics[width=1\linewidth]{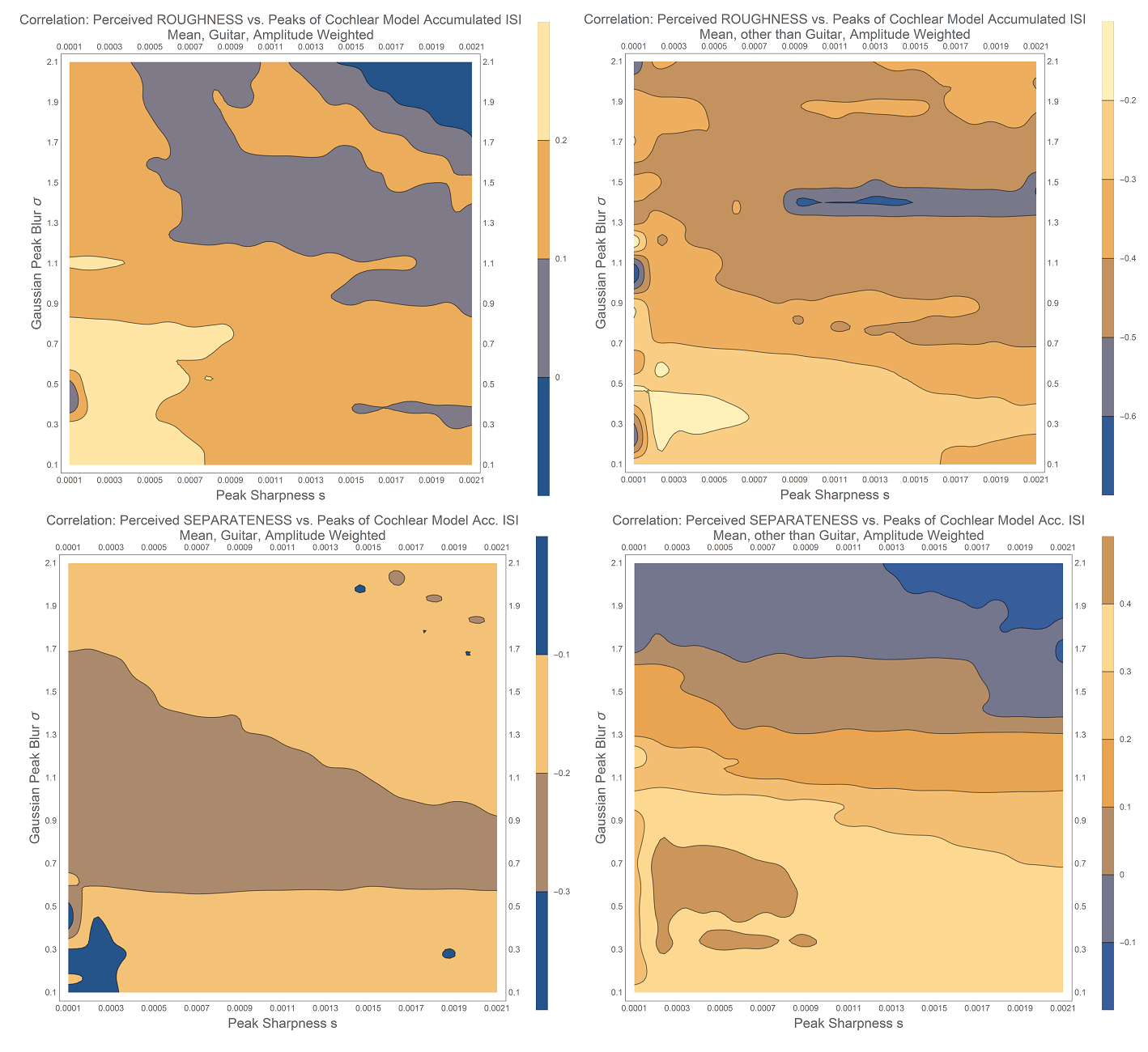}
		\caption{Same as Fig. \ref{fig:correntropymean} but correlation over amplitude-weighted ISI histogram W. The plots are considerably different from the entropy S Fig. \ref{fig:correntropymean} and number of amplitudes N in Fig. \ref{fig:corrmean}. Although the guitar roughness plot on the left top shows a kind of ridge as seen before, the correlations are very small. The roughness non-guitar plot on the right top has large negative correlations at a line at $\sigma \sim 1.4$. The separateness guitar plot has again low correlations, while the non-guitar separateness plot has larger positive correlations in its lower part, slightly similar to the roughness non-guitar plot. Still it seems that amplitude-weighting is not reasonably correlating with perception.}
		\label{fig:corramp}
	\end{figure}
	
\begin{figure}[H]
		\centering
		\includegraphics[width=1\linewidth]{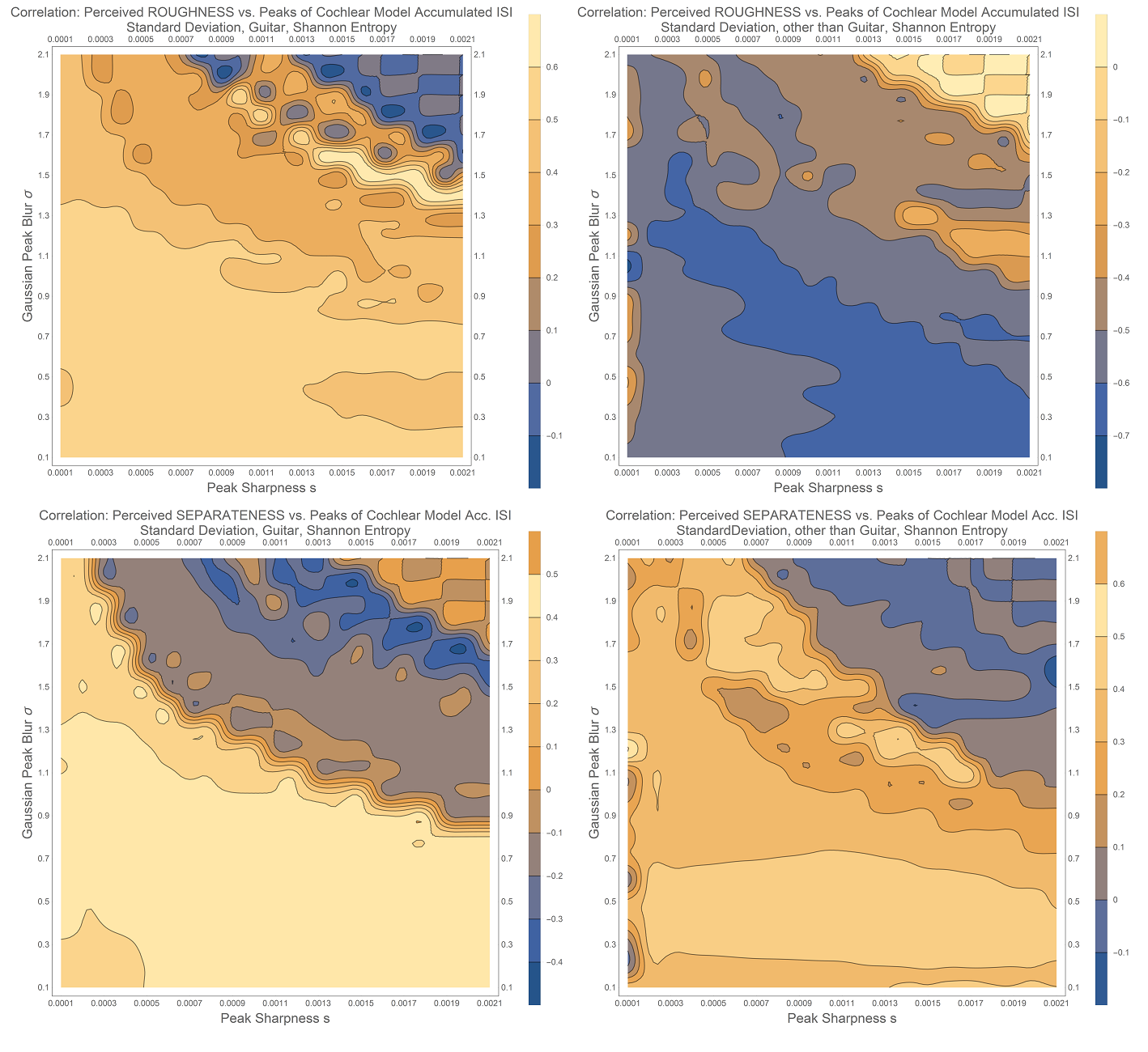}
		\caption{Correlation of the standard deviation of the perceptual separateness and roughness with the ISI histogram entropy S. The correlation shows slight similarity with the mean value correlations of the perceptual dimensions.}
		\label{fig:correntropysd}
	\end{figure}

	Finally in Fig. \ref{fig:correntropysd} the standard deviations of the perceptual data with the ISI histogram entropy S is shown. It has some similarities with the plots of the mean of the perceptual parameters. Roughness for the guitar sounds has again a ridge at the upper right corner, still the lower right side of the plot is slightly positive, not negative like with the mean values. Roughness for the non-guitar sounds is all negative, like the mean case. For separateness the guitar sounds have a negative ridge and a positive lower left corner, contrary to the mean values, this contrary behaviour is also seen with the non-guitar sounds compared to the mean case.

So overall a high standard deviation correlates positive with high entropy values in the lower left corner and negative with a low entropy in the upper right corner. Therefore we can conclude that when subjects were concentrating on strong coincidence detection, their judgments were much more consistent intersubjectively compared to the cases where subjects put their attention to low coincidence. The presence of only 3-5 amplitudes left after strong coincidence detection is pointing to a pitch perception subjects performed, while low coincidence detection can be associated with timbre. Therefore high correlations of perception with much coincidence, the ridges at theright upper corners of the plots, means that subjects had put their attention to pitch. Contrary, a high correlation of perception with low coincidence detection, the left lower part of the plots, indicates subjects attention to timbre. This is expected as pitch is a much more consistent perception than timbre.
	
\begin{figure}
		\centering
		\includegraphics[width=1\linewidth]{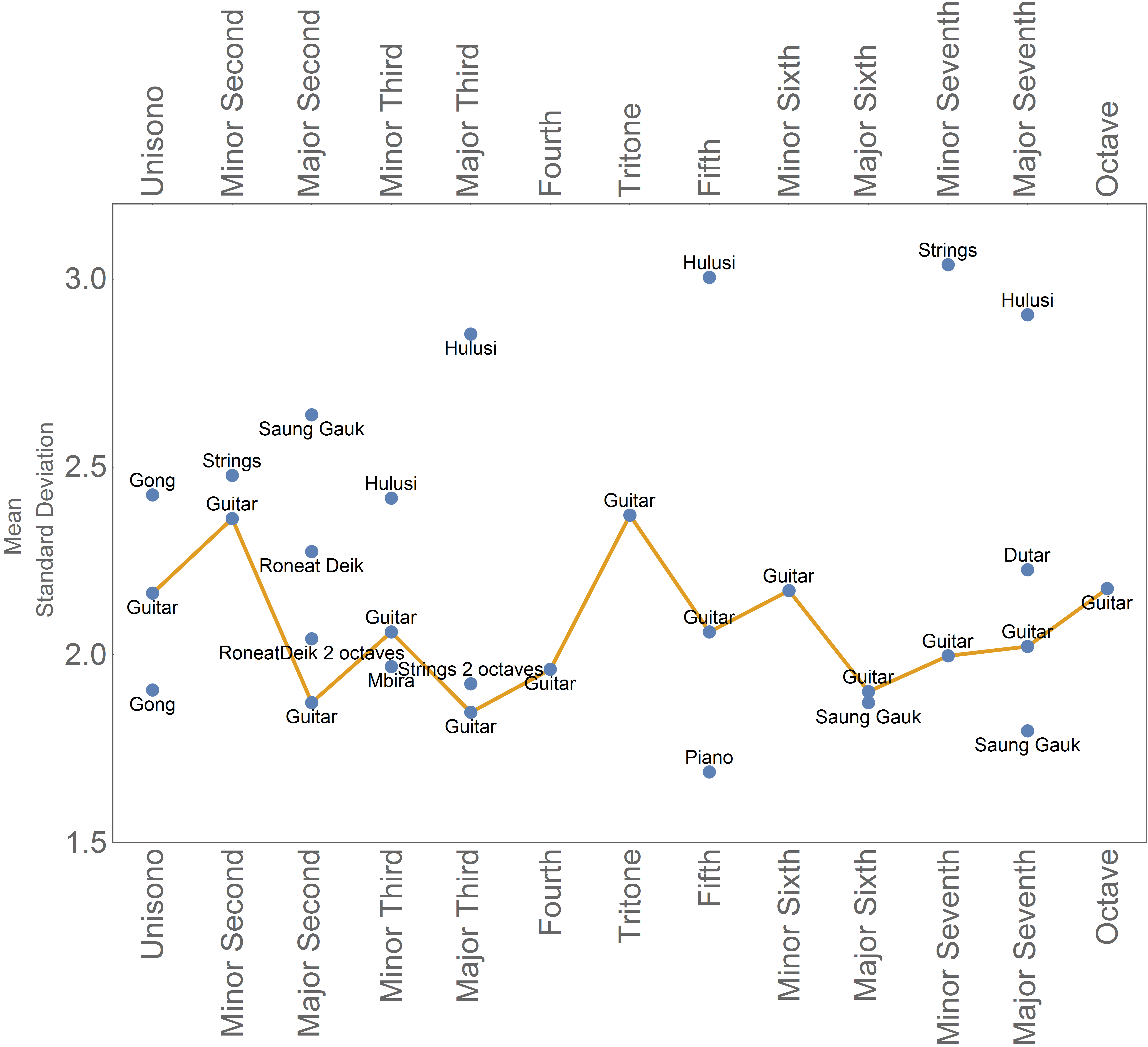}
		\caption{Standard deviation for roughness perception. The guitar tones are among those sounds with lowest standard deviation. Three \emph{hulusi} and one strings sound are considerably larger in their SD.}
		\label{fig:perceptionroughnessstandarddeviation}
	\end{figure}

\begin{figure}
		\centering
		\includegraphics[width=1\linewidth]{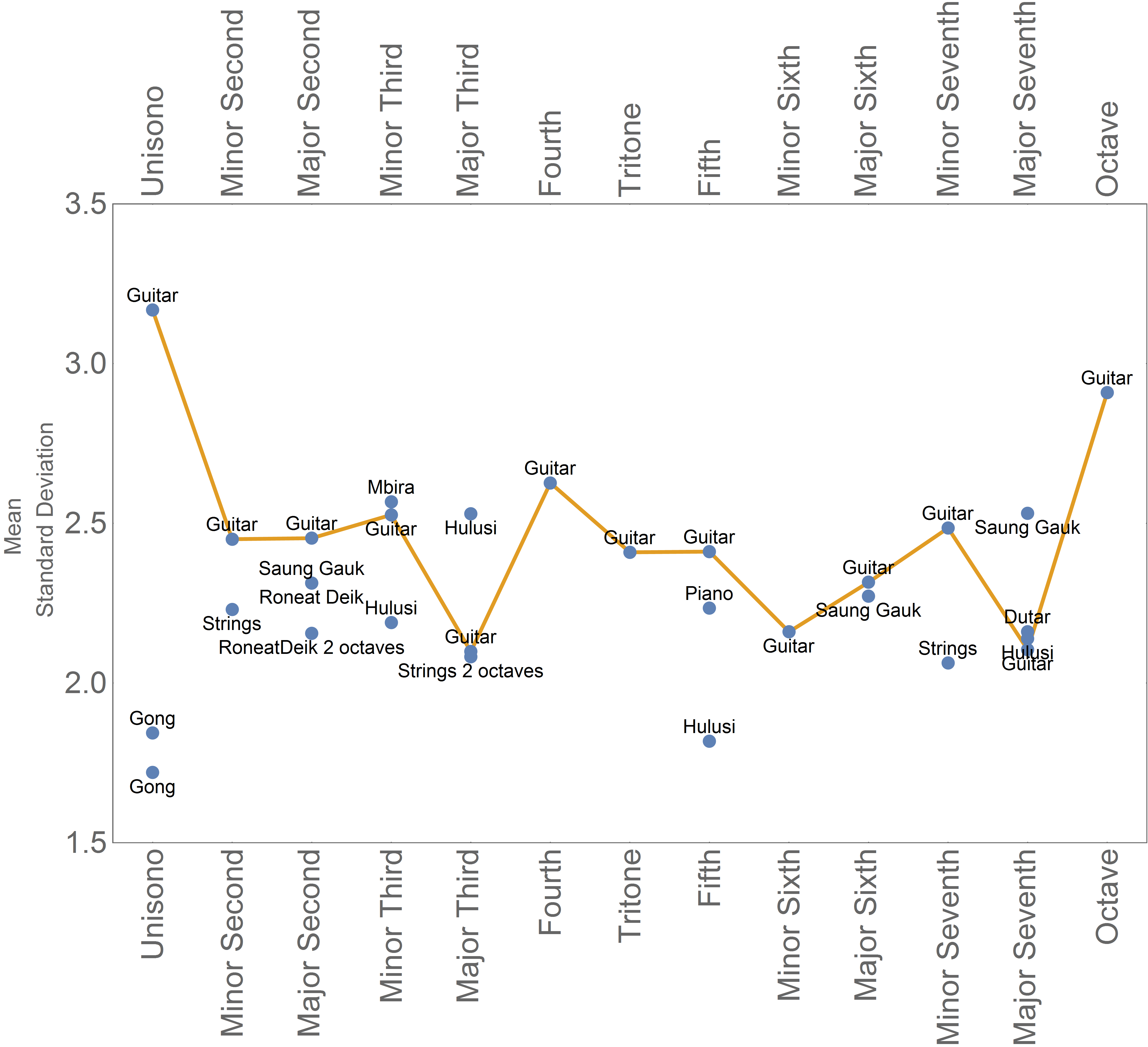}
		\caption{Perceptual standard deviation of the separateness sounds. Here the guitar tones are among those sounds with the highest SD. The gongs and one \emph{hulusi} have considerably lower values.}
		\label{fig:perceptionseparatenessstandarddeviation}
	\end{figure}

	It is also interesting to have a look at the standard deviations for the perception of roughness and separateness shown in Fig. \ref{fig:perceptionroughnessstandarddeviation} and Fig. \ref{fig:perceptionseparatenessstandarddeviation} respectively. For roughness the guitar sounds have among the lowest values, while for separateness they have the highest compared to the non-guitar sounds. The correlation between mean and standard deviation of roughness perception is 0.62, pointing to increased uncertainty of judgments with higher values. Still the correlation between mean and standard deviation for separateness is only 0.27 and so seem to be more independent. 
	
	So we conclude three points.
	
	First, the familiar guitar sounds are perceived using only a small amount of ISI histogram amplitudes, here 3-5, while the unfamiliar non-guitar tones are perceived using the broad range of many amplitudes in the ISI histogram. So the familiar sounds are perceived after intense coincidence detection, while the non-familiar sounds are perceived before coincidence detection, or only using a small part of it. As the main peaks of the ISI histogram are associated with pitch and the smaller amplitudes with timbre, we can conclude that perception strategy of familiar sounds is using pitch information and that of unfamiliar sounds using timbre.
	
	Second, separateness is correlated opposite between familiar (negative correlation) and unfamiliar (positive correlation) sounds with low coincidence detection (lower right plot corners), while roughness is correlated negative in both cases. This corresponds to the findings above, that separateness is perceived considerably stronger with guitar tones than with non-guitar sounds, while roughness does not show this clear distinction. So although for both it holds the pitch/timbre difference, separateness uses opposite strategies of perception while roughness uses the same correlation direction. 
	
	Third, judgments based on timbre are much more inconsistent than those based on pitch, which appears in the correlation between the standard deviation of the perceptual parameters and the entropy S of the ISI histograms. 
	
	These findings are based on the entropy calculated from the post-processed ISI histogram. Fig. \ref{fig:corrmean} shows the same correlations only taking the number N of surviving amplitudes into consideration. The results are similar in some respect, but show differences too. With the roughness guitar plot the ridge is again there, with negative correlation in the right left corner as before. Still now the unfamiliar roughness case has a strong negative correlation when few amplitudes remain in the upper right corner and strong positive correlations in the many amplitude case in the lower left corner. Again the perception is opposite, but much clearer than in the entropy case.
	
	For separateness the cases reverses too, the opposite correlation in the lower left corner disappear, in the familiar and unfamiliar case both regions correlate positive. Here the ridge region changing correlation direction.
	
	These findings contradict the findings from the perception test, namely the clear split of familiar and unfamiliar sounds for separateness, pointing to an opposite perception strategy. Therefore we conclude that using only amplitude counting is not perfectly in correspondence with the the listening test and therefore more unlikely to be used by subjects to judge separateness.
	
	Finally the third method of weighted amplitudes W is shown in Fig.\ref{fig:corramp}. Here the smallest correlations occur. Also the clear structures which have been found with entropy S and number of amplitudes N does not appear. Only the roughness guitar plot has some similarity with the ridge structure seen before. Therefore we conclude that the amplitude weighted ISI histogram is not correlating with perception.
	
	\section{Conclusions}
	
	The results point to two different perception strategies when listeners are asked to judge separateness and roughness of familiar and unfamiliar sounds. With familiar sounds a strong coincidence detection seems to take place, where listeners concentrate on the pitch extraction of sounds. With unfamiliar sounds they use the raw input much stronger without the coincidence detection reduction, concentrating on timbre. This corresponds to the overall experience with these sounds listeners reported for the unfamiliar sounds, namely that they could often not tell if the sounds consisted of two or only one pitch, or could not clearly identify one of the pitches. Therefore they needed to rely on timbre rather than on extracted pitches.
	
	A second perception strategy seem to happen when listeners are asked to judge separateness or roughness of sounds. With roughness timbre is the part listeners concentrate on, with separateness perception the strategy change between familiar and unfamiliar sounds. With unfamiliar sounds separateness is correlated positive with a complex sound, with familiar sounds timbre is correlated negative. So in unfamiliar multi-pitch sounds the pitches are perceived separate when the timbre is complex, while with familiar sounds the pitches are found separate when the timbre is more simple. It might be that listeners have is easier with familiar sounds to separate pitches when the timbre is more simple. Still if they are not familiar with the sound and not able to extract pitches anyway, they find complex spectra more separate than fused.
	
	In all cases judgments based on timbre are much more inconsistent between subjects than those based on pitch.
	
	In terms of neural correlates of roughness and separateness perception, the entropy of the ISI histogram is much better able to explain perception compared to the number of amplitudes or, even worse, the weighted amplitude sum. This is expected, as entropy is a more complex way of summing a sensation into a single judgment or separateness or roughness than the other methods, as entropy gives an estimation over the distribution of amplitudes in the spectrum and therefore their relation. Other methods might even be suited better, which is beyond the scope of the present paper to discuss.
	
	Overall it appears that many judgments of subjects are based on low-level parts of the auditory pathway before extended coincidence detection. This points to the necessity of perceiving timbre and pitch as a field of neural activity.

	\newpage
	\addtocounter{page}{2}

\end{document}